\documentclass[journal,12pt]{IEEEtran}
\usepackage{amssymb}
\usepackage{amsmath}
\usepackage{subcaption}
\usepackage{graphicx}
\usepackage{cite} 

\begin{document}

\title{Assessing the Viability of Fresnel Lenses for Weed Control in Prickly Pear Cactus Cultivation: A Spatiotemporal Coverage Perspective}

\author{Euzeli~C.~dos~Santos~Jr.,~\IEEEmembership{Senior Member,~IEEE,}
        Josinaldo~Lopes~Araujo~Rocha,
        Anielson~dos~Santos~Souza,
        Isaac Soares de Freitas,
        and Hudson~E.~Alencar~Menezes
\thanks{E. C. dos Santos Jr. is with Purdue University, Indianapolis, IN 46077 USA (e-mail: edossant@purdue.edu).}%
\thanks{J. L. A. Rocha, A. dos S. Souza, H. E. A. Menezes are with the Federal University of Campina Grande, Campina Grande, Paraíba, Brazil.}%
\thanks{I. S. de Freitas is with the Federal University of Paraíba, João Pessoa, Paraíba, Brazil.}
}

\maketitle

\begin{abstract}
In tropical semi-arid regions, prickly pear cactus (Opuntia ficus-indica) has emerged as a vital forage resource due to its high drought tolerance and minimal water requirements. However, even limited weed infestation can severely compromise cactus productivity, as the species is highly sensitive to competition for essential resources, which includes water, mineral nutrients, and sun exposure. Conventional herbicide-based weed control strategies face growing limitations due to resistance development and environmental concerns, underscoring the need for sustainable alternatives. This study revisits the historically underexplored application of linear Fresnel lenses (LFLs) for thermal weed control and establishes the technical feasibility of a contemporary autonomous weed management system that incorporates LFL technology within an unmanned ground vehicle (UGV) platform. Leveraging real-time image processing, georeferencing, and mechanical actuation, the system can perform a two-phase operation—weed mapping during non-optimal solar hours and targeted solar termination during peak irradiance. Analytical modeling quantifies the effective area coverage and time constraints imposed by the solar window. Preliminary results indicate that, while unsuitable for dense weed infestations, the system presents a viable solution for precision, post-emergent weed control in sparse infestations. The favorable solar geometry in tropical zones, especially in the Brazilian semiarid region, and the targeted nature of the approach make this technology particularly well-suited for sustainable agriculture in under-resourced regions.
\end{abstract}

\begin{IEEEkeywords}
Precision agriculture, Linear Fresnel lens, solar thermal weed control, tropical agriculture.
\end{IEEEkeywords}

\section{Introduction}

In regions with semi-arid climates in tropical zones, where drought can restrict the availability of conventional forage options \cite{amar2022}, incorporating prickly pear cactus as a supplementary feed for cattle proves to be highly beneficial \cite{sipango2022}. The advantages are found in the plant's ability to grow without irrigation and its high production rate, making it an optimal selection for dry areas. Although not providing a complete nutrition source on its own, the prickly pear cactus can furnish energy, water, and fiber, assisting cattle in not only surviving but also thriving during dry periods. Effective weed control in Opuntia ficus-indica (prickly pear cactus) cultivation is essential, as unmanaged weed growth competes aggressively for water, light, and soil nutrients, thereby impeding cactus development, reducing forage yield, and compromising overall crop productivity in semi-arid systems \cite{araya1998}.

Indeed, weed proliferation remains one of the most critical challenges in modern agriculture, often leading to significant yield losses and increased competition for water, nutrients, and light. While chemical herbicides have historically been the dominant method for weed suppression, growing concerns over herbicide resistance, environmental degradation, and regulatory restrictions have spurred the search for sustainable, non-chemical alternatives \cite{boyd2022}.

\begin{figure}[h]
\centering
\includegraphics[width=0.8\linewidth]{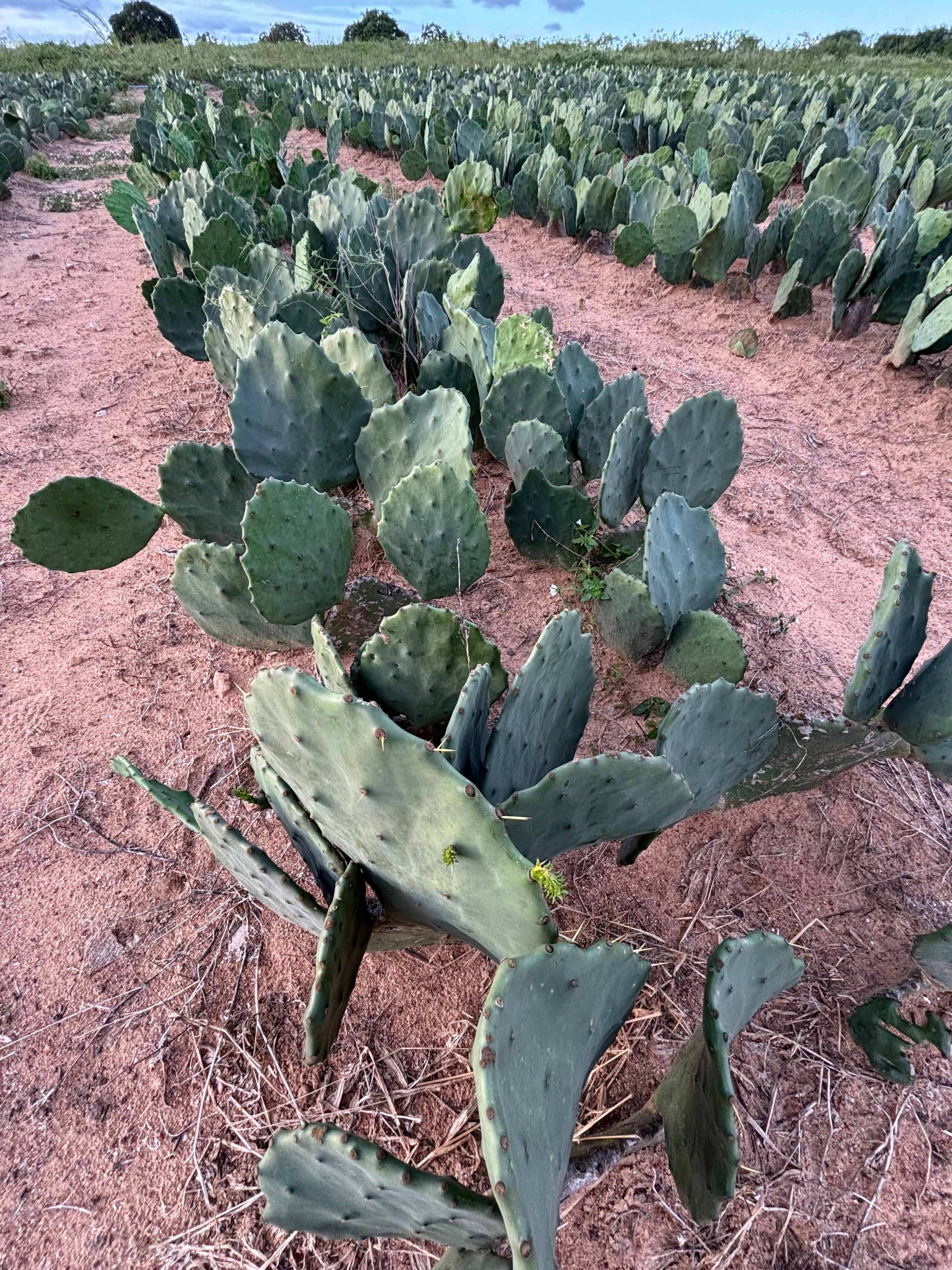}
\caption{Opuntia ficus-indica (prickly pear cactus) cultivation site commonly utilized as a forage resource for cattle in semi-arid areas.}
\label{fig:field}
\end{figure}

Among these alternatives, the use of concentrated solar energy through Fresnel lens systems has received periodic attention over the past several decades. Early experimental studies demonstrated the efficacy of linear Fresnel lenses in delivering lethal doses of solar thermal energy to both weed seedlings and seeds. The work done by \cite{johnson1989seeds} showed that exposure of ten different weed species to concentrated solar radiation using a curved linear Fresnel lens resulted in 100\% seed mortality for several species within 20 seconds of exposure, provided that the seeds were on or near the soil surface. In a complementary study, \cite{johnson1989plants} demonstrated that the application of linear Fresnel lens–focused sunlight to weed seedlings, such as redroot pigweed, kochia, and green foxtail, resulted in high mortality within a few seconds of treatment, with effectiveness dependent on species, developmental stage, and soil moisture conditions.

Despite these promising results, Fresnel lens–based weed control has not been widely adopted commercially. Its limitations (including dependence on solar irradiance, narrow focal zones, and lack of integration with mechanized farming systems) have constrained its scalability and practical utility. However, recent advancements in precision agriculture, including autonomous ground vehicles, real-time sensing, and GPS-guided actuation, have opened new opportunities to revisit this approach under a more controlled and targeted framework.

Patent literature has also acknowledged the potential of such technologies. For instance, \cite{uspatent2009} proposes the use of Fresnel lenses mounted on human-operated mobile platforms to selectively irradiate weeds with concentrated sunlight. The concept introduces a method for field-scale weed control using solar energy, but it lacks integration with modern autonomous technologies and adaptive dosing mechanisms.

The present work builds on these earlier foundations and introduces a novel, fully autonomous unmanned ground vehicle (UGV) designed to deploy linear Fresnel lenses for chemical-free, site-specific weed management. By leveraging advancements in sensor technologies, machine vision, and robotic actuation, the proposed system is capable of detecting weed presence, aligning the focal axis, and delivering a precise thermal “dose” of solar energy sufficient to induce irreversible cellular damage in the target plant. This method obviates the reliance on chemical inputs, presenting a sustainable weed management strategy that aligns with the principles of modern precision agriculture, particularly suited to tropical regions. However, developing nations within tropical zones frequently adopt technologies originating from more economically and technologically advanced countries in temperate and subtropical regions. This incongruity arises from financial constraints and disparate levels of technological maturation. Furthermore, the geographical positioning of these tropical nations, characterized by unique solar radiation patterns, often renders imported technologies less than optimally effective, highlighting a significant disconnect between adopted methods and environmental realities.

\begin{figure}[t]
\centering
\includegraphics[width=0.9\linewidth]{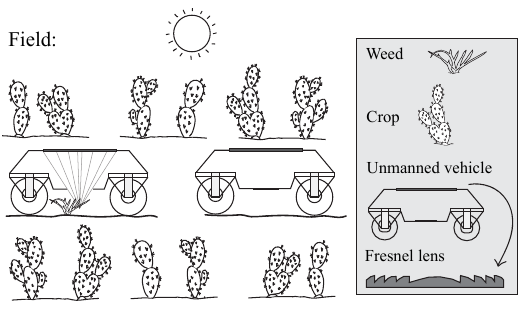}
\caption{Weed management system integrating an unmanned ground vehicle equipped with a linear Fresnel lens for targeted solar concentration.}
\label{fig:field}
\end{figure}

The proposed weed management system, illustrated in Fig.~\ref{fig:field}, consists of an unmanned ground vehicle (UGV) equipped with a linear Fresnel lens and an embedded sensing system designed for autonomous navigation and real-time weed detection. The UGV traverses the crop field autonomously, utilizing onboard cameras and electronic sensors to monitor the inter-row space for the presence of weeds. In the absence of weed detection, a mechanical shading mechanism remains engaged to block direct sunlight from reaching the Fresnel lens, thereby preventing unnecessary energy concentration and potential damage to surrounding crops or soil. However, when a weed is identified along the crop line, the system automatically actuates the optical path to expose the Fresnel lens to sunlight. The lens then focuses the solar radiation onto the target weed, delivering a localized and chemical-free thermal dose sufficient to induce tissue damage and plant desiccation. This selective activation of the Fresnel lens ensures energy efficiency and precision in weed control while preserving crop integrity. This paper further investigates, through a mathematical and analytical modeling approach, the effective area that can be treated using this method, given the duration of solar exposure required to lethally affect specific weed species. The analysis also demonstrates that, when combined with modern advances in autonomous sensing, actuation, and vehicle control, the Fresnel-lens-based system presents a technically feasible and economically viable solution for selective weed management in tropical agricultural zones. These regions, characterized by high solar incidence and year-round crop production, offer an ideal environment for revitalizing and deploying solar-based weed control technologies.

\section{Tropical Zone Characteristics and Thermal Weed Control}

\subsection{Solar Geometry in Tropical Zone}

Tropical zones, defined between the Tropic of Cancer (23.45°N) and the Tropic of Capricorn (23.45°S), receive high solar irradiance due to the high sun angle throughout the year. Countries such as Brazil, Nigeria, and regions of North Africa (e.g., Egypt, Sudan) benefit from nearly zenithal solar incidence around noon for much of the year, particularly near the equator. This solar geometry is ideal for optical systems such as linear Fresnel lens, which concentrate direct beam radiation.

\begin{figure}[h]
\centering
\includegraphics[width=1.0\linewidth]{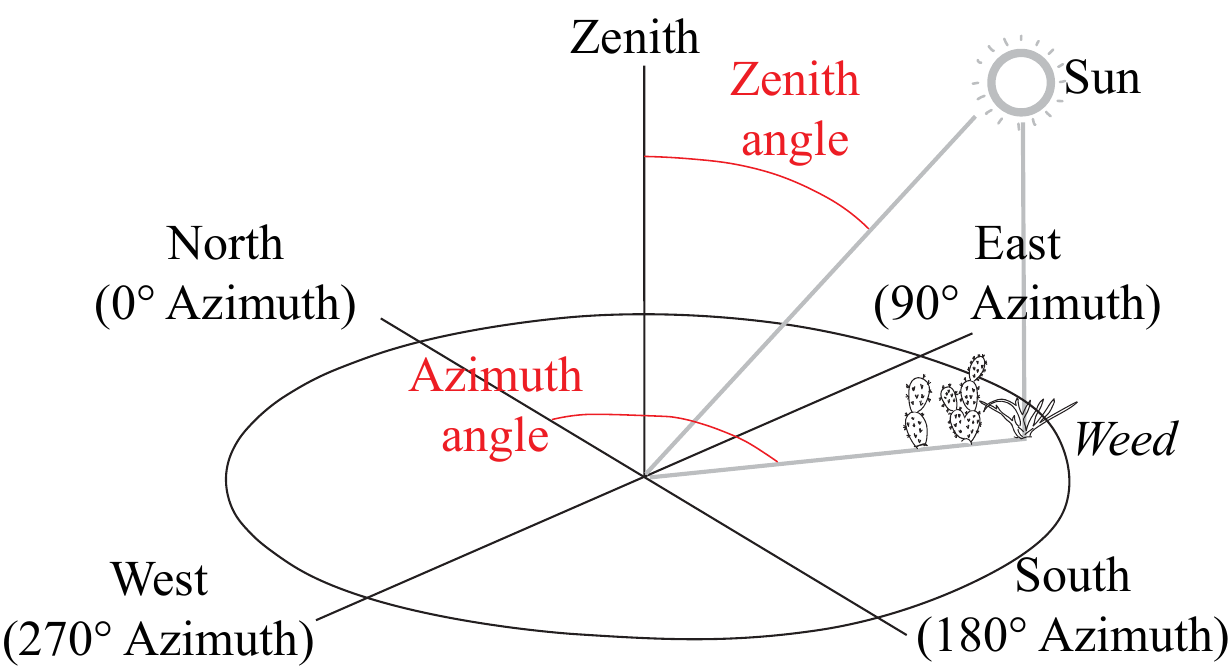}
\caption{Schematic representation of solar geometry with solar zenith angle and solar azimuth angle highlighted in red.}
\label{fig:angles}
\end{figure}

In these regions the solar altitude angle (elevation above the horizon) can approach 90° (i.e., directly overhead) at noon. The solar declination oscillates between -23.45° and +23.45° annually, aligning closely with the latitude of these regions during solstices. The solar position is described using two angles: (i) solar zenith angle ($\theta_z$), i.e., angle between the sun's rays and the vertical, and (ii) solar azimuth angle, which compasses direction from which sunlight originates, as presented in Fig. \ref{fig:angles}.  These parameters are fundamental to understanding the high-angle solar incidence typical of tropical regions, which enables efficient use of horizontal concentrators such as linear Fresnel lenses.

The zenith angle is calculated as:
\begin{equation}
    \cos(\theta_z) = \sin(\phi)\sin(\delta) + \cos(\phi)\cos(\delta)\cos(h)
\end{equation}
where, $\phi$ is the latitude, $\delta$ is the solar declination angle, $h$ is the hour angle, defined as $h = 15^\circ \times (\text{solar time} - 12)$.

At solar noon, $h = 0$, thus:
\begin{equation}
    \theta_{z,\text{noon}} = \arccos(\sin(\phi)\sin(\delta) + \cos(\phi)\cos(\delta))
\end{equation}

For instance, at $\phi = 0^\circ$ (e.g., equator) and $\delta = 0^\circ$ (equinox), $\theta_z = 0^\circ$. Also, at $\phi = 10^\circ$ and $\delta = 10^\circ$, $\theta_z = 0^\circ$ at solar noon. Thus, in many tropical locations, sunlight reaches the Earth's surface at near-perpendicular angles during midday, ideal for Fresnel lenses oriented horizontally.

\subsection{Principle of Thermal Weed Control}

Thermal weed control is based on the principle that thermal energy induces lethal damage to plant tissues~\cite{oliveira2018}. Lethal temperatures typically range between 55\,\textdegree{}C and 70\,\textdegree{}C and are applied directly to meristematic or foliar tissues for a few seconds~\cite{dahlquist2007,jacobs2024}. This method becomes particularly advantageous in palm cultivation due to its targeted action, which significantly reduces risks to non-target plants~\cite{ansley2007}.

At the physiological and biochemical levels, heat denatures essential proteins and enzymes, ruptures cell membranes---causing loss of integrity and solute leakage---and coagulates the cytoplasm~\cite{kristoffersen2008}. From a photosynthetic perspective, excessive heat irreversibly inactivates photosystem~II (PSII), degrades chlorophyll, and disrupts thylakoid structures, thereby inhibiting photosynthesis~\cite{taiz2017}. Cellular turgor loss and structural collapse are immediate consequences~\cite{kristoffersen2008}. As a result, plant death occurs through the synergistic effects of these injuries, with the apical meristem and delicate tissues (e.g., the hypocotyl) exhibiting the highest sensitivity~\cite{taiz2017}.

The efficacy of thermal weed control depends on several factors, including tissue moisture content (hydrated plants are more susceptible), cuticle thickness, developmental stage (younger plants are more vulnerable), exposure duration, and weed species~\cite{merfield2017,bauer2020}. This method not only eliminates emerged weeds but also inhibits seed germination~\cite{dahlquist2007,jacobs2024}.

\section{Use of Fresnel Lens}

\begin{figure}[h]
\centering
\begin{subfigure}[b]{0.8\linewidth}
    \centering
    \includegraphics[width=\linewidth]{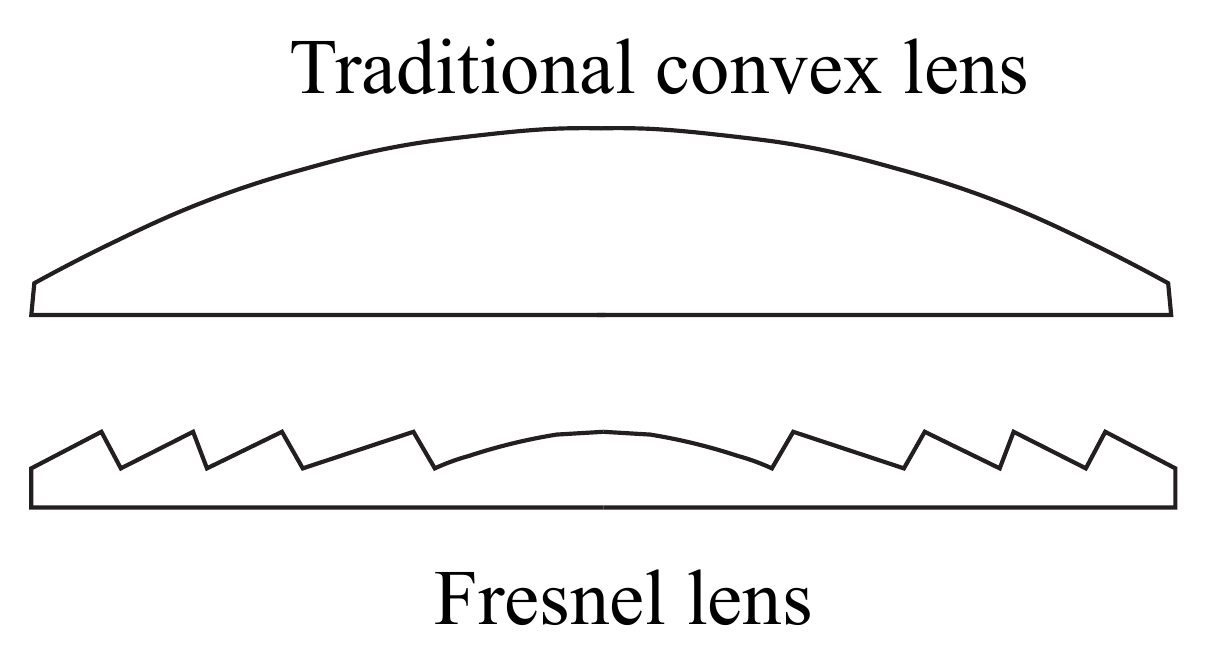}
    \label{fig:subfig1}
\end{subfigure}
\hfill
\begin{subfigure}[b]{1.0\linewidth}
    \centering
    \includegraphics[width=\linewidth]{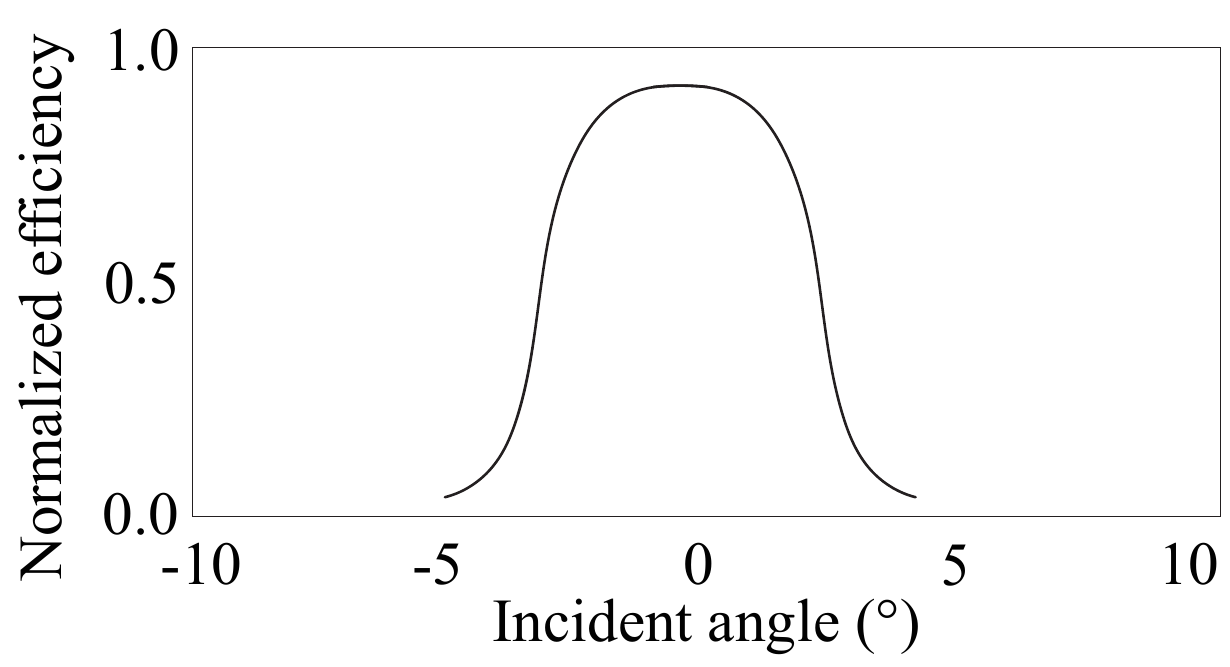}
    \label{fig:subfig2}
\end{subfigure}
\caption{(top) Comparison between traditional convex and Fresnel lens. (bottom) Normalized efficiency versus incident angle.}
\label{fig:fresnel}
\end{figure}

Fresnel lenses are lightweight, portable optical elements widely employed for light concentration and magnification applications. The original concept was introduced by the French physicist Augustin-Jean Fresnel, who utilized this design to construct the first glass Fresnel lens for lighthouse illumination in 1822 \cite{yadav2024}. Unlike conventional spherical or aspherical lenses, which rely on continuous curvature, Fresnel lenses consist of a series of narrow, concentric grooves inscribed onto one surface of a flat substrate, typically composed of lightweight polymeric material. This groove structure replicates the focusing properties of a conventional lens while significantly reducing the overall thickness, mass, and manufacturing cost, see Fig. \ref{fig:fresnel} (top).

The optical efficiency $\eta$ of a Linear Fresnel Lens (LFL) system depends on several angle-dependent parameters and can be expressed as:
\begin{equation}
    \eta(\theta_i) = T(\theta_i) \cdot C_g(\theta_i) \cdot \cos(\theta_i)
\end{equation}
In this expression, $T(\theta_i)$ represents the transmittance of the lens as a function of the solar incidence angle $\theta_i$, $C_g(\theta_i)$ denotes the geometric concentration ratio at the same angle, and $\cos(\theta_i)$ accounts for the projection of solar irradiance onto the lens aperture due to the angular position of the sun. A typical normalized efficiency versus incident angle is presented in Fig. \ref{fig:fresnel} (bottom) \cite{smith2021}. The maximal optical efficiency happens at zero degrees, where the incident rays are normal to the lens surface. In this condition, transmittance losses due to Fresnel reflection are minimized, the geometric concentration ratio is at its peak, and the projection of irradiance onto the lens aperture is maximal.

As the solar incidence angle increases in magnitude, three compounding effects reduce the overall optical efficiency. First, transmittance $T(\theta_i)$ decreases due to the growing reflectance at the air–lens interface, as described by the Fresnel equations for unpolarized light. Second, the geometric concentration ratio $C_g(\theta_i)$ degrades with increasing $\theta_i$ because the focal line becomes blurred and elongated, thereby reducing the energy density at the receiver plane. Third, the projection factor $\cos(\theta_i)$ decreases as the angle between the solar rays and the lens normal increases, which further diminishes the effective irradiance entering the lens aperture.

\begin{figure}[t]
\centering
\includegraphics[width=1.0\linewidth]{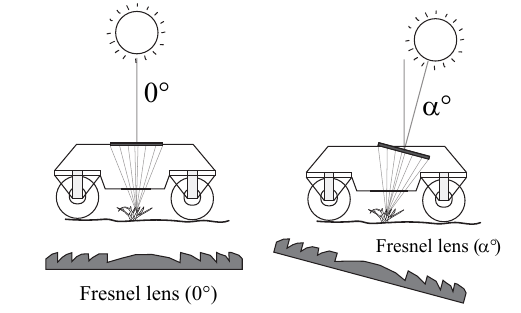}
\caption{Fresnel lens mechanical control.}
\label{fig:sun_angles}
\end{figure}

During solar noon in tropical zones, the incidence angle approaches zero, i.e., $\theta_i \approx 0^\circ$, and thus $\cos(\theta_i) \approx 1$. Under this condition, reflection losses are minimized, leading to transmittance values that approximate $T(\theta_i) \approx T_0$. Simultaneously, the focal spot is sharply defined and well-aligned with the receiver, which results in a maximized geometric concentration ratio, i.e., $C_g(\theta_i)$ approaches its theoretical maximum. Since solar irradiance is nearly perpendicular to the lens surface at this time, the projection factor remains close to unity. Indeed, the efficiency shown in Fig. \ref{fig:fresnel} (bottom) suggests that the lenses need to in parallel with the solar rays, which will be compenrated automatically as presented in Fig. \ref{fig:sun_angles}. 

\section{Daily Solar Window}

At solar noon, when $h = 0$, the zenith angle $\theta_{z,\text{noon}}$ attains its minimum value, which corresponds to the sun's highest elevation in the sky for a given day. To determine the time interval during which the zenith angle remains below a predefined threshold $\theta_{\text{max}}$ (e.g., $25^\circ$), (1) can be rearranged and solved for the hour angle $h$ that satisfies the equality $\theta_z = \theta_{\text{max}}$. Substituting this into the zenith angle equation yields

\begin{equation}
    \cos(\theta_{\text{max}}) = \sin(\phi)\sin(\delta) + \cos(\phi)\cos(\delta)\cos(h).
\end{equation}

Isolating the cosine term leads to

\begin{equation}
    \cos(h) = \frac{\cos(\theta_{\text{max}}) - \sin(\phi)\sin(\delta)}{\cos(\phi)\cos(\delta)}.
\label{eq:cosh}
\end{equation}

Once $\cos(h)$ is computed, the corresponding hour angle bounds at which the solar zenith angle reaches $\theta_{\text{max}}$ are given by

\begin{equation}
    h = \pm \arccos\left( \frac{\cos(\theta_{\text{max}}) - \sin(\phi)\sin(\delta)}{\cos(\phi)\cos(\delta)} \right).
\label{eq:h_angle}
\end{equation}

To express this interval in units of time, it is noted that the hour angle progresses at a rate of $15^\circ$ per hour. Thus, the time interval on either side of solar noon during which the zenith angle remains below the specified threshold is

\begin{equation}
    t = \frac{1}{15} \cdot \arccos\left( \frac{\cos(\theta_{\text{max}}) - \sin(\phi)\sin(\delta)}{\cos(\phi)\cos(\delta)} \right),
\end{equation}
where $t$ is in hours. Therefore, the total duration $\Delta t$ during which the sun remains within the zenith angle constraint is given by

\begin{equation}
    \Delta t = \frac{2}{15} \cdot \arccos\left( \frac{\cos(\theta_{\text{max}}) - \sin(\phi)\sin(\delta)}{\cos(\phi)\cos(\delta)} \right).
    \label{eq:duration}
\end{equation}

Equation~\eqref{eq:duration} provides a generalized expression for the number of daylight hours per day during which the sun remains within an angular distance $\theta_{\text{max}}$ of the zenith, as a function of geographic latitude $\phi$ and solar declination $\delta$. This expression is valid for all values of $\phi$ and $\delta$ for which the argument of the arccosine function lies within the interval $[-1, 1]$, ensuring that the sun does, at some point, approach the specified zenith angle. This formulation is particularly relevant for fixed-orientation solar concentrators such as linear Fresnel lenses, whose optical efficiency is strongly dependent on the angular proximity of incident solar rays to the surface normal.

\begin{figure}[h]
\centering
\includegraphics[width=1.0\linewidth]{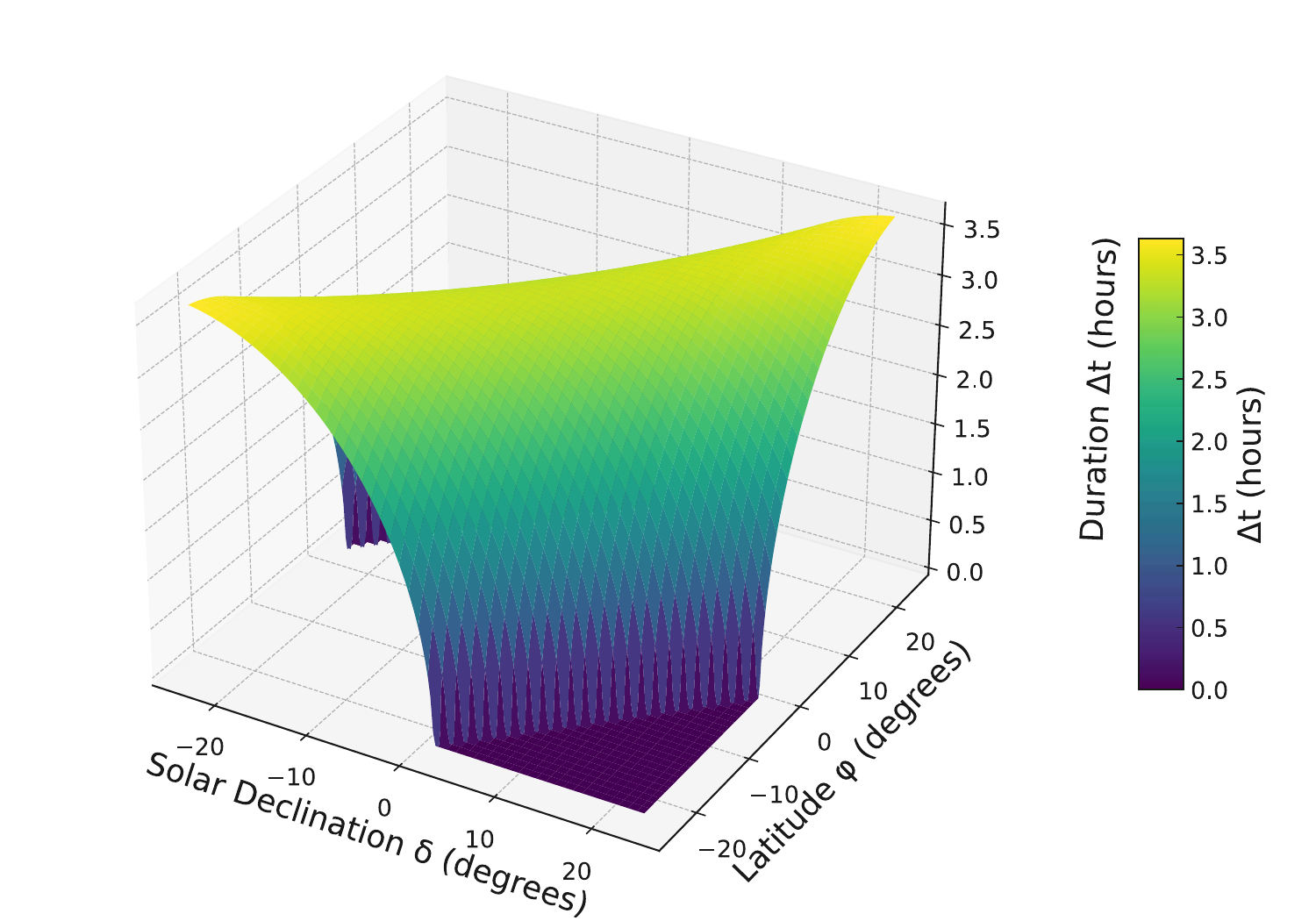}
\caption{Variation of daily duration $\Delta t$ versus solar declination and latitude.}
\label{fig:zenith_duration}
\end{figure}

Fig.~\ref{fig:zenith_duration} illustrates the variation of daily duration $\Delta t$, expressed in hours, during which the solar zenith angle remains within $25^\circ$ of the vertical. The duration is presented as a function of geographic latitude $\phi$ and solar declination $\delta$, both ranging from $-23.45^\circ$ to $+23.45^\circ$, which defines the bounds of the tropical zone. The results confirm that the maximum duration occurs when the solar declination closely matches the geographic latitude, i.e., when $\delta \approx \phi$. Under this condition, the sun passes nearly overhead at solar noon, and the zenith angle remains below the specified threshold for the longest portion of the day, reaching up to approximately 3.34 hours. As the difference between the solar declination and the site latitude increases, the total time within the high-angle solar window rapidly decreases. This angular sensitivity underscores the importance of geographic positioning and seasonal timing for the effective operation of solar concentrators such as linear Fresnel lenses. The analysis also demonstrates that high-angle solar availability is inherently dynamic, driven by both the Earth's axial tilt and orbital progression, and therefore must be accounted for when assessing the temporal viability of solar thermal applications in tropical agricultural systems. As the latitude and/or solar declination changes, the mechanical tracking system seen in Fig. \ref{fig:sun_angles} will compensate that variation. 

\section{Autonomous Deployment Justification}

The consistent and favorable solar geometry in tropical regions supports the deployment of an autonomous Unmanned Ground Vehicle (UGV) equipped with Linear Fresnel Lenses. Such a system can operate optimally during the time window of 3.5h, when solar irradiance is high and the incidence angle is such that can be compensated by $\alpha$. The travel speed of the UGV can be modulated dynamically to control the duration of exposure applied to each weed target. 

\begin{figure*}[h]
\centering
\includegraphics[width=0.8\linewidth]{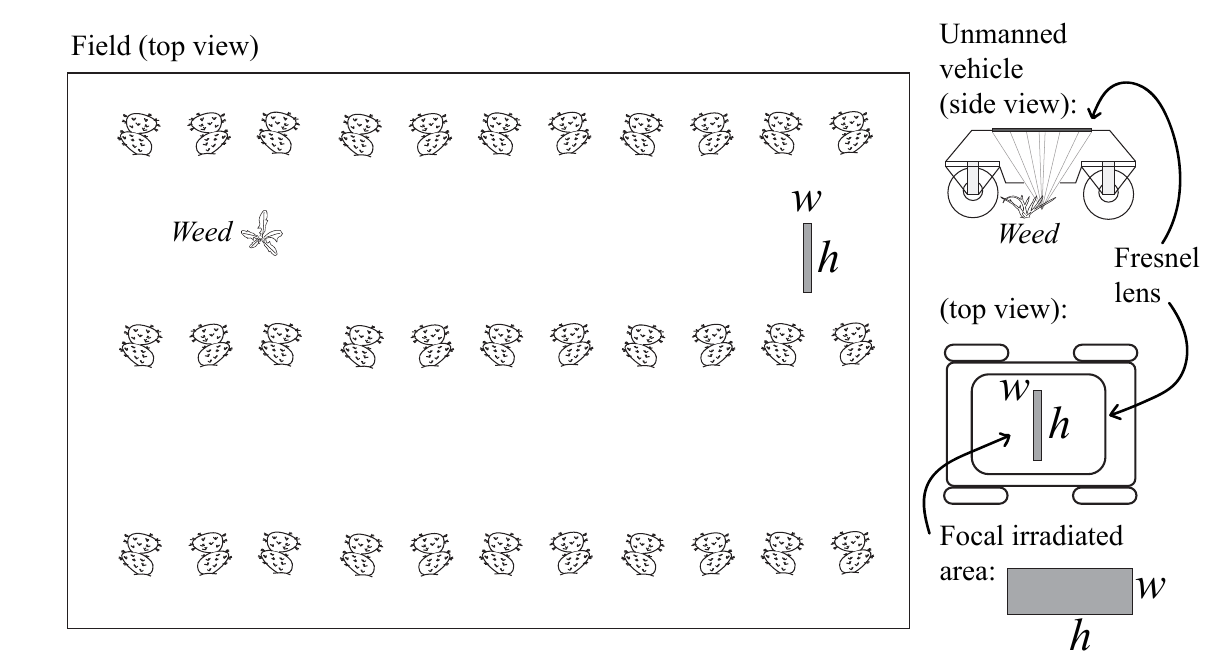}
\caption{Top view of the focal irradiated area where concentrated solar energy is applied for targeted weed control.}
\label{fig:topview}
\end{figure*}

To determine the total area covered by a moving system over a period of 3.5h, we begin with the known parameters. Let the cross-sectional area being swept be denoted by \( A = w \times h \) (footprint affected by the solar concentrator system - see Fig. \ref{fig:topview}), where \( w \) and \( h \) represent the width and height, respectively. If this area $A$ is translated linearly along a given direction with constant speed $v$, then over a time interval $t$, the total area swept by the system is given by

\begin{equation}
    A_{\text{swept}}(t) = h \cdot (v \cdot t),
\end{equation}

which simplifies to

\begin{equation}
    A_{\text{swept}}(t) = A \cdot \frac{v \cdot t}{w}.
\end{equation}

Assuming a target total area $A_{\text{total}}$ must be completely covered by the moving system, the required cumulative operating time $t_{\text{total}}$ to accomplish this task satisfies

\begin{equation}
    A_{\text{total}} = A \cdot \frac{v \cdot t_{\text{total}}}{w},
\end{equation}

which can be rearranged to yield

\begin{equation}
    t_{\text{total}} = \frac{A_{\text{total}} \cdot w}{A \cdot v} = \frac{A_{\text{total}}}{h \cdot v}.
\end{equation}

\begin{figure}[h]
\centering
\includegraphics[width=1.0\linewidth]{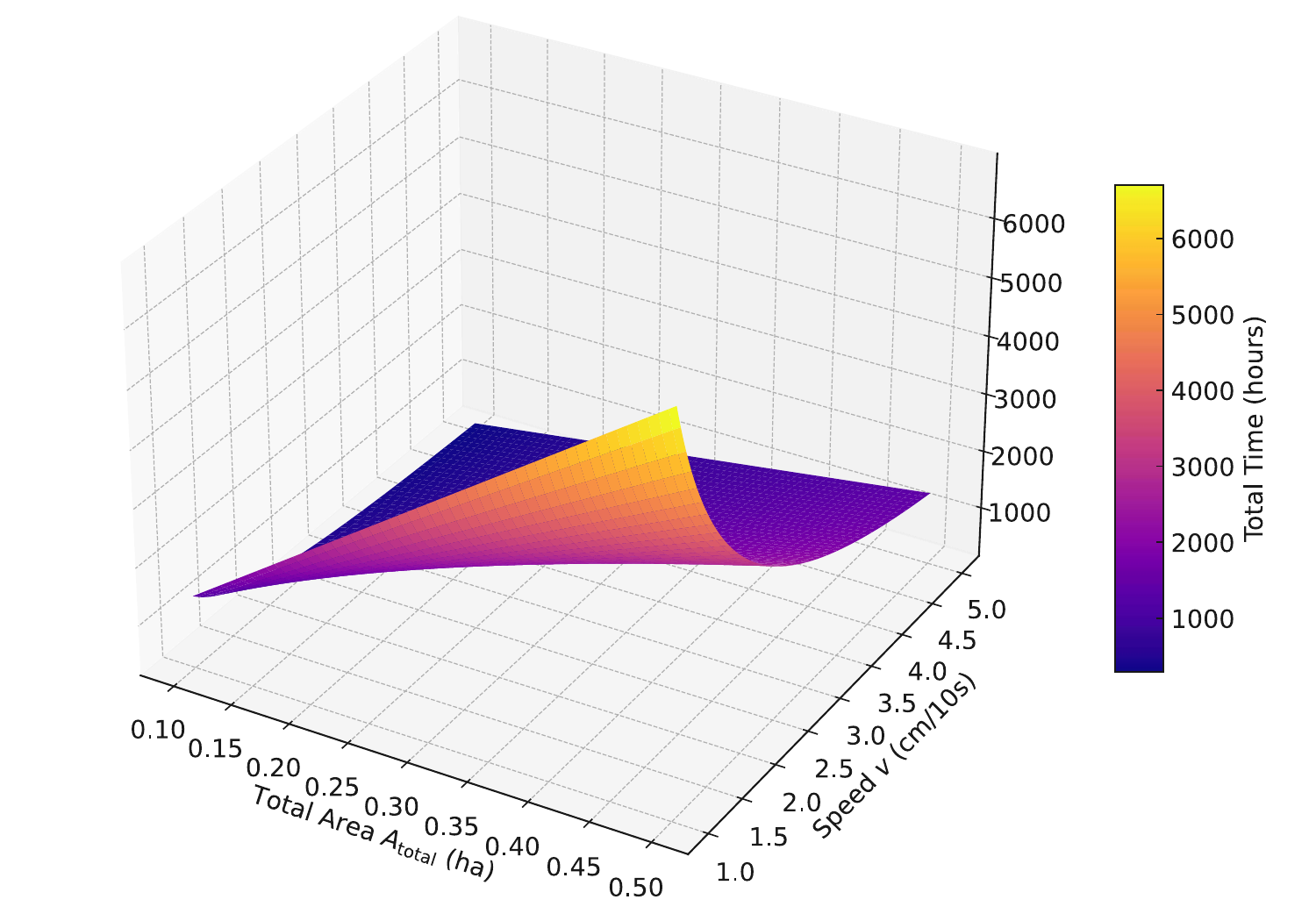}
\caption{Total time versus speed and area for effective weed control.}
\label{fig:time3D}
\end{figure}

This expression quantifies the total time required for the system to traverse and treat the entire area $A_{\text{total}}$ based on its geometrical and kinematic parameters. Fig. \ref{fig:time3D} shows the total time versus the speed of application and the total area. The speed interval was chosen based on the minimum exposure duration required to achieve effective weed termination, ensuring sufficient thermal or optical dosage. Meanwhile, the target area range, spanning from 0.1ha to 0.5ha, reflects typical operational scales encountered in precision agriculture applications. Given that the system operates solely under conditions of direct or near-direct solar irradiance and is constrained to a limited daily operational window of $T_{\text{day}} = 3.5$ hours, the total number of hours required to cover even a relatively small area renders its practical application infeasible.

Let $D$ represent the total number of operational days required to complete the full coverage. Then the number of days is determined by dividing the total time by the number of hours available per day:

\begin{equation}
    D = \frac{t_{\text{total}}}{T_{\text{day}}} = \frac{A_{\text{total}}}{h \cdot v \cdot T_{\text{day}}}.
\end{equation}

This formulation provides an analytical estimate of the total number of days required for a solar-powered, linearly moving concentrator system to cover a prescribed area $A_{\text{total}}$ under restricted daily operating conditions. 

\section{Practical Use of Fresnel Lenses in Cactus Cultivation}

Based on the preceding analysis, it becomes evident that the deployment of a commercially available concentrated solar device is not a practical solution for weed control in scenarios where the field is heavily infested. However, when integrated into an autonomous platform capable of detecting individual weed emergence and targeting them selectively, such a device demonstrates significant potential as an effective and sustainable tool for site-specific weed management.

\begin{figure}[h]
\centering
\includegraphics[width=1.0\linewidth]{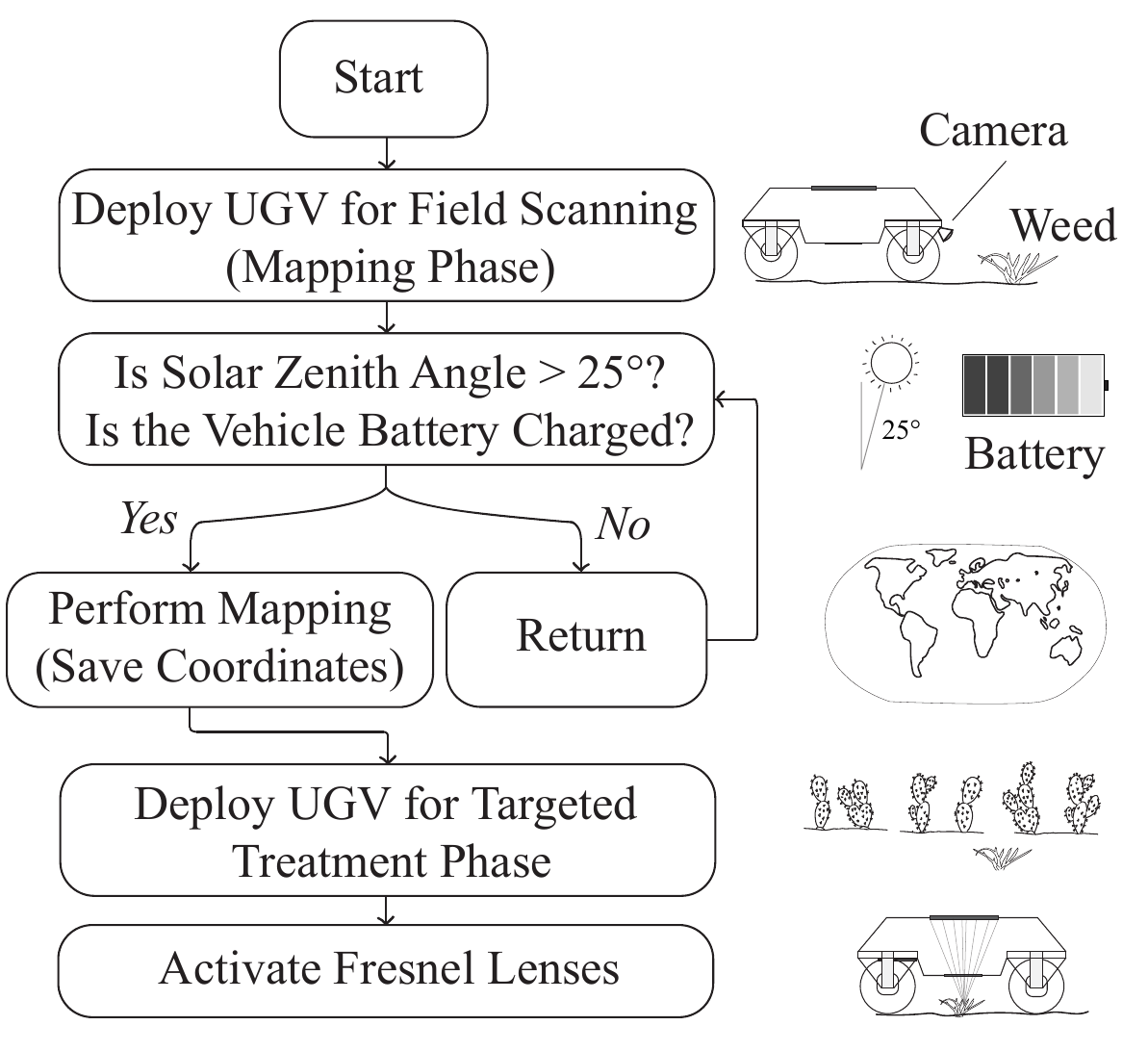}
\caption{Conceptual flow chart of the proposed autonomous weed management system employing concentrated solar energy.}
\label{fig:flow}
\end{figure}

The proposed management system is designed to operate autonomously across the target field, as shown in the flow chart of Fig. \ref{fig:flow}, which performs a two-phase process to enable efficient and localized weed control. In the initial phase, the unmanned ground vehicle (UGV) traverses the entire area while employing an embedded vision system composed of cameras and real-time image processing algorithms. This system enables the identification and georeferencing of weed-infested locations with high spatial resolution. This initial mapping phase is conducted during periods when the solar zenith angle exceeds $25^\circ$, corresponding to times of day that are unsuitable for weed treatment due to suboptimal solar incidence conditions.  

Following the completion of the detection and mapping stage, the UGV is redeployed with the explicit objective of executing targeted weed eradication. During this second phase, only the previously identified locations are addressed, thereby optimizing energy consumption and operational time. This approach allows for precise intervention in post-emergence scenarios, where weeds tend to reappear in spatially heterogeneous patterns, enhancing the feasibility and sustainability of the weed management process.

Prickly pear cactus (Opuntia ficus-indica) constitutes an exceptionally suitable candidate for the application of site-specific, solar-based weed control technologies, particularly due to its widespread cultivation in semi-arid regions within the tropical zone. In these environments, where annual precipitation is markedly low, the proliferation of weeds tends to be naturally constrained in comparison to more humid climates. Nonetheless, even minimal weed emergence can exert disproportionately negative effects on prickly pear crops, as this species exhibits low tolerance to interspecific competition for essential resources such as water, nutrients, and sunlight. Consequently, the implementation of a selective and non-chemical weed management strategy is not only contextually appropriate but also crucial to safeguarding the productivity and sustainability of prickly pear cultivation systems.

\section{Conclution}

This work demonstrates the feasibility of employing linear Fresnel lenses for chemical-free, targeted weed control in tropical semi-arid environments through integration with autonomous ground platforms. The high solar elevation angles characteristic of tropical zones provide a compelling justification for the use of solar concentrators, enabling efficient thermal delivery during the hours close to midday. Although conventional applications of Fresnel lenses have been limited by angular sensitivity and lack of mechanization, recent advances in robotics, computer vision, and energy-efficient actuation allow for precise and selective deployment of solar thermal energy. The proposed system’s dual-phase operational mapping ensures both energy efficiency and agronomic precision. Mathematical modeling confirms that large-scale applications remain constrained by the narrow solar window and slow operational speeds. However, for localized or post-emergent weed outbreaks, especially in crops such as prickly pear cactus that are intolerant to competition, this technology offers a sustainable and scalable solution. Ultimately, this study revives and modernizes a historically overlooked approach, aligning it with the contemporary goals of precision agriculture, sustainability, and regional adaptability.

\bibliographystyle{IEEEtran}

\begin{thebibliography}{99}

\bibitem{amar2022}
A.~L.~Amar, R.~Congdon, R.~Coventry, and C.~Gardiner, ``Responses of selected tropical forage legumes to imposed drought,'' \textit{Science Archives}, vol.~3, no.~3, pp.~158--167, 2022.

\bibitem{sipango2022}
N.~Sipango, K.~Khuliso, E.~Ravhuhali et~al., ``Prickly pear (Opuntia spp.) as an invasive species and a potential fodder resource for ruminant animals,'' \textit{Sustainability}, vol.~14, no.~7, article 3719, 2022.

\bibitem{araya1998}
H.~J.~Araya and W.~W.~Smith, ``Forage productivity of \textit{Opuntia ficus-indica} under different management practices in Ethiopia,'' \textit{J. Arid Environ.}, vol.~38, no.~3, pp.~333--345, 1998.

\bibitem{boyd2022}
N.~S. Boyd and R.~F. Warfield, “Occurrence and management of herbicide resistance in annual vegetable production,” *Weed Sci.*, vol. 70, no. 5, pp. 586–598, 2022.

\bibitem{smith2021}
J.~Smith and M.~Kumar, ``Design of Novel Compound Fresnel Lens for High-PerformancePhotovoltaic Concentrator,'' \textit{Appl. Opt. Eng.}, vol.~58, no.~4, pp.~345--352, 2021.

\bibitem{yadav2024}
S.~Yadav, P.~Jain, and P.~Singh, ``Solar energy concentrator research: past and present,'' \textit{Solar Energy Concentrators: Essentials and Applications}, pp.~121--136, 2024.






\bibitem{johnson1989seeds}
B. J. Johnson, J. J. Jank, and J. R. Brown, ``Response of Seed of 10 Weed Species to Fresnel-lens-concentrated Solar Radiation,'' \textit{Weed Technology}, vol. 4, no. 1, pp. 157--162, 1990.

\bibitem{johnson1989plants}
B. J. Johnson, J. R. Brown, and J. J. Jank, ``Response of Monocot and Dicot Weed Species to Fresnel Lens Concentrated Solar Radiation,'' \textit{Weed Science}, vol. 37, no. 5, pp. 651--657, 1989.

\bibitem{uspatent2009}
J. K. Morehead, ``Method and Apparatus for Controlling Weeds with Solar Energy,'' U.S. Patent Application US20090114210A1, May 2009.


\bibitem{ansley2007}
R.~J.~Ansley and M.~J.~Castellano, ``Prickly Pear Cactus Responses to Summer and Winter Fires,'' \textit{Rangeland Ecology \& Management}, vol.~60, no.~3, pp.~244--252, 2007.

\bibitem{bauer2020}
T.~Bauer, C.~Lindner, and H.~Wichmann, ``Thermal weed control: Mechanisms and impact on seed germination,'' \textit{Weed Science}, vol.~68, no.~3, pp.~245--253, 2020.

\bibitem{dahlquist2007}
R.~M.~Dahlquist, C.~M.~Prather, and J.~J.~Stark, ``Time and temperature requirements for weed seed thermal death,'' \textit{Weed Science}, vol.~55, no.~5, pp.~619--625, 2007.

\bibitem{jacobs2024}
A.~Jacobs, L.~Nguyen, and P.~Meyer, ``Advanced thermal techniques for sustainable weed management,'' \textit{Crop Protection}, vol.~176, pp.~106--114, 2024.

\bibitem{kristoffersen2008}
P.~Kristoffersen and S.~U.~Larsen, ``Non-chemical weed control on traffic islands: Efficacy of five techniques,'' \textit{Weed Research}, vol.~48, no.~2, pp.~124--130, 2008.

\bibitem{merfield2017}
C.~N.~Merfield, J.~P.~Travaille, and D.~G.~Saville, ``Relative susceptibility of weeds to heat treatment,'' \textit{Biological Agriculture \& Horticulture}, vol.~33, no.~4, pp.~197--206, 2017.

\bibitem{oliveira2018}
M.~F.~Oliveira and A.~M.~Brighenti, ``Thermal weed control: Principles and applications,'' in \textit{Non-Chemical Weed Management}, Springer, pp.~89--112, 2018.

\bibitem{taiz2017}
L.~Taiz, E.~Zeiger, I.~Møller, and A.~Murphy, \textit{Plant Physiology and Development}, 6th ed., Sunderland, MA: Sinauer Associates, 2017. (Chapter 12: Stress Physiology).


\end{thebibliography}

\end{document}